\pdfoutput=1

\documentclass[11pt]{article}

\usepackage{naacl2021}

\usepackage{times}
\usepackage{latexsym}

\usepackage[T1]{fontenc}

\usepackage[utf8]{inputenc}

\usepackage{microtype}

\usepackage{booktabs}
\usepackage{graphicx}
\usepackage{xcolor}
\usepackage{enumitem}
\usepackage{amssymb}
\usepackage{multirow}
\usepackage{hologo}
\usepackage{mdframed}
\usepackage[frozencache=true,cachedir=.]{minted}

\title{Redefining Absent Keyphrases and their Effect on Retrieval Effectiveness}

\author{Florian Boudin \\
  LS2N, Université de Nantes, France \\
  \normalsize{\texttt{florian.boudin@univ-nantes.fr}} \\\And
  Ygor Gallina \\
  LS2N, Université de Nantes, France \\
  \normalsize{\texttt{ygor.gallina@univ-nantes.fr}} \\}
  
\begin{document}

\maketitle

\begin{abstract}
Neural keyphrase generation models have recently attracted much interest due to their ability to output \emph{absent keyphrases}, that is, keyphrases that do not appear in the source text.
In this paper, we discuss the usefulness of absent keyphrases from an Information Retrieval (IR) perspective, and show that the commonly drawn distinction between present and absent keyphrases is not made explicit enough.
We introduce a finer-grained categorization scheme that sheds more light on the impact of absent keyphrases on scientific document retrieval.
Under this scheme, we find that only a fraction (around 20\%) of the words that make up keyphrases actually serves as document expansion, but that this small fraction of words is behind much of the gains observed in retrieval effectiveness. 
%
We also discuss how the proposed scheme can offer a new angle to evaluate the output of neural keyphrase generation models.

%
%
%
%
\end{abstract}

\section{Introduction}

Searching the scholarly literature for documents of interest is becoming frustratingly difficult and time-consuming as the volume of published research grows exponentially.
%
%
One promising approach to address this problem and improve the retrievability of documents is to supplement paper indexing with automatically generated keyphrases~\citep{zhai-1997-fast,10.5555/338985.338996,boudin-etal-2020-keyphrase}.
Traditionally, keyphrases are defined as a short list of terms that represent the main concepts in a document~\cite{10.1023/A:1009976227802}.
In recent years, this definition was further refined to differentiate between keyphrases that are present in the source document or not, and in turn, proposed models for producing keyphrases were divided into extractive~\citep[\textit{inter alia}]{florescu-caragea-2017-positionrank,boudin-2018-unsupervised,10.1145/3331184.3331219,wang-etal-2020-incorporating,santosh-etal-2020-sasake} and generative models~\citep[\textit{inter alia}]{meng-etal-2017-deep,zhao-zhang-2019-incorporating,chen-etal-2020-exclusive,bahuleyan-el-asri-2020-diverse} based on their ability to output absent keyphrases.
%

Obviously, keyphrases have different effects on retrieval models depending on whether or not they occur in the document: \emph{present keyphrases} highlight important parts of the input and make weighting terms easier, while \emph{absent keyphrases} add new terms to the input and provide some form of document expansion.
Intuitively, assigning absent keyphrases is more appealing since it may alleviate the \emph{vocabulary mismatch} problem between query terms and relevant documents~\cite{10.1145/32206.32212}, hence enabling the retrieval of relevant documents that otherwise would have been missed. 
This is especially true for scholarly collections, in which documents are mostly short texts (i.e.~scientific abstracts) due to licensing issues and/or resource limitations~\cite{10.1145/3295750.3298953}.
Yet, the extent to which present and absent keyphrases contribute to improved retrieval effectiveness has not been thoroughly explored.
Worse still, there is no unique and rigorous definition of what exactly makes a keyphrase absent.

\newcommand{\reducedstrut}{\vrule width 0pt height .9\ht\strutbox depth .9\dp\strutbox\relax}
\newcommand{\hl}[2]{%
  \begingroup
  \setlength{\fboxsep}{0pt}%
  \colorbox{#1}{\reducedstrut#2\/}%
  \endgroup
}

\colorlet{c1}{blue!30!white}
\colorlet{c2}{red!30!white}
\colorlet{c3}{purple!30!white}
\colorlet{c4}{green!35!white}
\colorlet{c5}{yellow!40!white}
\colorlet{c6}{orange!30!white}

\begin{figure*}[ht]

\begin{mdframed}[backgroundcolor=blue!1] 
\textbf{Study on the Structure of Index Data for \hl{c1}{Metasearch} \hl{c3}{System}} \\[-.6em]

This paper proposes a new technique for \hl{c1}{Metasearch} \hl{c3}{system}, which is based on the grouping of both keywords and URLs. This technique enables \hl{c1}{metasearch} \hl{c3}{systems} to \hl{c5}{share} \hl{c4}{information} and to reflect the estimation of \hl{c6}{users'} preference. With this \hl{c3}{system}, \hl{c6}{users} can search not only by their own keywords but by similarity of HTML documents. In this paper, we describe the principle of the grouping technique as well as the summary of the existing \hl{c2}{search} \hl{c3}{systems}. \\[-.6em]

\textbf{Present kps}:
\hl{c1}{Metasearch} --
\hl{c2}{Search} \hl{c3}{System} \\[-.8em]

\textbf{Absent kps}:
\hl{c4}{Information} \hl{c5}{Sharing} --
\hl{c4}{Information} Retrieval --
\hl{c6}{User}'s Behavior --
Retrieval Support \\[-1.4em]


{\small
\textcolor{gray}{
\hspace{1.9cm} \hspace{.9cm} \underline{R}eordered \hspace{.9cm}
\hspace{0.1cm} \hspace{1.3cm} \underline{M}ixed \hspace{1.3cm}
\hspace{0.1cm} \hspace{.8cm} \underline{M}ixed \hspace{.8cm}
\hspace{0.1cm} \hspace{.8cm} \underline{U}nseen \hspace{.8cm}
}
}
\end{mdframed}

\caption{Sample document (title, abstract) from the NTCIR-2 test collection (docid: gakkai-e-0001384947). Author-assigned keyphrases are divided into present and absent using token-level matching with stemming. Finer-grained categories for absent keyphrases (i.e.~\underline{R}eordered, \underline{M}ixed and \underline{U}nseen) are also outlined.}
\label{fig:example}
\end{figure*}

Although not stated explicitly, many recent studies adopt the definition by~\cite{meng-etal-2017-deep}, in which keyphrases that do not match any contiguous subsequence of source text are regarded as absent.
From an Information Retrieval (IR) perspective where stemmed content words are used to index documents, this definition is not sufficiently explicit, as demonstrated by the example shown in Figure~\ref{fig:example}.
We see that, under this definition, some absent keyphrases can have all of their words occurring in the source document, and therefore act no differently from present keyphrases on indexing.
In fact, only a fraction of the words that compose these absent keyphrases are genuinely expanding the document, which in our example are the set of words $\lceil$\emph{retrieval}, \emph{behavior}, \emph{support}$\rfloor$.
From a keyphrase generation point of view, this definition is not entirely satisfactory either, since training a model to produce absent keyphrases from an output vocabulary, while some of these might actually be reconstructed from the source document, is arguably overkill.
Here, we argue that this may be one reason behind the poor performance of current sequence-to-sequence models in generating absent keyphrases~\cite{10.1145/3383583.3398517}.


In this paper, we advocate for a stricter definition of absent keyphrases and propose a fine-grained categorization scheme that reflects how many new words are introduced within each keyphrase.
Through this scheme, we shed new light on the effect of absent keyphrases on document retrieval effectiveness, and provide insights as to why current models for keyphrase generation are unable to accurately produce absent keyphrases.
As a by-product, we introduce a new benchmark dataset for scientific document retrieval through the task of context-aware citation recommendation, that is composed of 169 manually extracted queries with relevance judgments and a collection of over 100K documents on topics related to IR.


\section{(Re)defining Absent Keyphrases}
\label{sec:def}

Telling absent and present keyphrases apart may seem quite easy at first, but actually there are several intricacies to the process that should be noted.
Starting from~\citet{meng-etal-2017-deep}'s definition, ``\emph{we denote phrases that do not match any contiguous subsequence of source text as absent keyphrases, and the ones that fully match a part of the text as present keyphrases}'', it is apparent that simple string matching between keyphrases and source document is not acceptable since it produces false positives (e.g.~``\emph{supervised learning}'' matches ``\emph{\textcolor{gray}{un}supervised learning}'').
Instead, token-level sequence matching is to be used and combined with stemming to deal with different inflectional forms of the same word.
Using stemming is critical here since it is carried out as a standard procedure in indexing documents for IR, but also in evaluating the precision of keyphrase generation models against gold standard annotations~\cite{hasan-ng-2014-automatic}.

Looking back at our example in Figure~\ref{fig:example}, we see that absent keyphrases can be further divided into three sub-categories depending on the proportion of present words they contain.
Indeed, some absent keyphrases have some, or even all, of their constituent words (in stemmed forms) present in the text, while others are composed entirely of unseen words.
Accordingly, we propose the following fine-grained categorization scheme (illustrated with the example from Figure~\ref{fig:example} and explained in more depth with pseudo-code in Appendix~\ref{sec:code}):

\begin{itemize}[wide=0.5em,leftmargin=0.5em,itemsep=0.05em,font=\bfseries]

    \item[\underline{P}resent:] keyphrases that match contiguous sequences of words in the source document (e.g.~``\emph{Search System}'').
    
    \item[\underline{R}eordered:] keyphrases whose constituent words occur in the source document but not as contiguous sequences (e.g.~``\emph{Information Sharing}'').
    
    \item[\underline{M}ixed:] keyphrases from which some, but not all, of their constituent words occur in the source document (e.g.~``\emph{Information Retrieval}'').
    
    \item[\underline{U}nseen:] keyphrases whose constituent words do not occur in the source document (e.g.~``\emph{Retrieval Support}'').
    
\end{itemize}

In contrast to the previously-used binary classification (i.e.~present or absent), this finer-grained categorization scheme draws a distinction between keyphrases that expand the document (i.e.~mixed and unseen) and those that don't (i.e.~present and reordered).
%
It thus allows us to better understand how keyphrases affect the retrieval process by making it possible to numerically quantify the contribution of each category to the overall retrieval effectiveness.
At the same time, this scheme provides a new angle to evaluate the ability of keyphrase generation models to output absent keyphrases by contrasting their PRMU distributions against those observed in the gold standard annotations.
In other words, a model has to mimic the distribution of absent keyphrases in manual annotation in order to perform well.


\section{Experiments}

Here, we outline our experimental setup (\S\ref{sec:exp-setup}), examine the distribution of keyphrases in commonly-used datasets with respect to the proposed categorization scheme (\S\ref{sec:dist}), show the influence of each category on the retrieval effectiveness (\S\ref{sec:ir}), and explore how these categories fit into the outputs of neural keyphrase generation models (\S\ref{sec:kg}).


\subsection{Experimental settings}
\label{sec:exp-setup}

Experiments in \textit{ad-hoc} document retrieval are carried out on the NTCIR-2 test collection~\cite{ntcir-2} which is, to our knowledge, the only available benchmark dataset for that task.
It includes 322,058 scientific abstracts in English annotated with author-assigned keyphrases (4.8 per doc.~on avg.), and 49 search topics (queries) with relevance judgments.
%
%
Documents cover a wide range of domains from pure science to humanities, although half of the documents are about computer science.

Given the rather limited size of the NTCIR-2 test collection, we conducted additional experiments in context-aware citation recommendation~\cite{10.1145/1772690.1772734} which is the task of retrieving citations (documents) for a given text (query).
Since no publicly available keyphrase-annotated collection exists for that task, we created one by collecting documents (\hologo{BibTeX} entries) from the ACM Digital Library. 
Our dataset contains 102,411 documents in English on topics related to IR\footnote{We use the SIGs IR, KDD, CHI, WEB and MOD sponsored conferences and journals as a means to filter documents.}, most of which (69.2\%) have author-assigned keyphrases (4.5 per doc.~on avg.).
We then followed the methodology proposed in~\cite{10.1145/3132847.3133085}, and selected 30 open-access scientific papers\footnote{Papers published in SIGIR, CHIIR, ICTIR or WSDM 2020 conferences.} from which we manually extracted 169 citation contexts (queries) and 481 cited references (relevant documents).
The resulting dataset, named ACM-CR, is publicly available\footnote{\url{https://github.com/boudinfl/redefining-absent-keyphrases/blob/main/data/acm-cr/acm-cr.v1.tar.gz}}.

For both retrieval tasks, we rank documents against queries using the standard BM25 model implemented in the Anserini\footnote{\url{http://anserini.io/}} open-source IR toolkit~\cite{Yang:2017:AEU:3077136.3080721}, on top of which we apply the RM3 query expansion technique~\cite{abdul2004umass} to achieve strong, near state-of-the-art retrieval results~\cite{Lin:2019:NHC:3308774.3308781,10.1145/3331184.3331340}.
For all models, we use Anserini’s default parameters.
We evaluate retrieval effectiveness in terms of mean average precision (mAP) on the top 1,000 retrieved documents for \textit{ad-hoc} document retrieval, and in terms of recall at 10 retrieved documents for context-aware citation recommendation as recommended in~\cite{farber2020citation}.
We use the Student’s paired t-test to assess statistical significance of our retrieval results at $p<0.05$~\cite{Smucker:2007:CSS:1321440.1321528}.

%

%


\begin{table}[!hbt]
    \centering
    \begin{tabular}{l|rrrr|r}
    \multicolumn{2}{c}{~} &
    \multicolumn{3}{c}{\small{$\lceil$ \hfill {\scriptsize absent keyphrases} \hfill $\rceil$}}
    \\
    \toprule
    \textbf{Dataset}  & \textbf{\%P} & \textbf{\%R} & \textbf{\%M} & \textbf{\%U} & \textbf{\%uw} \\
    \midrule
     NTCIR-2 & 61.9 &  8.1 & 16.5 & 13.5 & 21.4 \\
     ACM-CR  & 53.6 & 11.7 & 19.3 & 15.4 & 25.5\\
     KP20k   & 60.2 &  9.5 & 15.4 & 15.0 & 22.3 \\
    \bottomrule
    \multicolumn{1}{c}{~} &
    \multicolumn{2}{c@{}}{\small{$\lfloor$ \hfill {\scriptsize term-weighting} \hfill $\rfloor$}} &
    \multicolumn{2}{c}{\small{$\lfloor$ \hfill {\scriptsize doc. expansion} \hfill $\rfloor$}}
    \end{tabular}
    \caption{Proportion of \underline{P}resent, \underline{R}eordered, \underline{M}ixed and \underline{U}nseen keyphrases in datasets. We also report the ratio of unique, unseen words in M+U keyphrases (\%uw).}
    \label{tab:dist}
\end{table}

\begin{table*}[ht!]
    \centering
    \begin{tabular}{l|ccc||ccc}
        \multicolumn{1}{c}{} & \multicolumn{3}{c}{$\lceil$ \hfill NTCIR-2 (mAP) \hfill $\rceil$} & \multicolumn{3}{c}{$\lceil$ \hfill ACM-CR (recall@10) \hfill $\rceil$}\\
        \toprule
        \textbf{index} & \textbf{BM25} & \textbf{+RM3} & \textbf{\#kp} & \textbf{BM25} & \textbf{+RM3} & \textbf{\#kp}\\
        \midrule
            title \& abstract   & 29.55 & 32.83 & - & 35.64 & 34.09 & - \\
        \cmidrule(lr){1-7}
            ~+~\underline{P}resent   & 30.74$^\dagger$ & 33.47 & 2.9 & 36.02 & 34.09 & 2.4 \\
            ~+~\underline{R}eordered & 29.79 & 33.48 & 0.4  & 35.43 & 33.40 & 0.5 \\ 
            ~+~\underline{M}ixed     & \textbf{30.80}$^\dagger$ & 33.85 & 0.8 & 36.22 & 33.41 & 0.9 \\ 
            ~+~\underline{U}nseen    & 29.67 & \textbf{33.94} & 0.7 & \textbf{36.24} & 33.78 & 0.8 \\ 
        \cmidrule(lr){1-7}
            ~+~Absent \scriptsize{(R+M+U)}    & 30.77$^\dagger$ & 34.87$^\dagger$ & 1.9 & 36.62 & 34.10 & 2.1 \\ 
        \cmidrule(lr){1-7}
            ~+~Highlight \scriptsize{(P+R)}   & 30.64$^\dagger$ & 33.82 & 3.3  & 35.82 & 32.36 & 2.9\\ 
            ~+~Expand \scriptsize{(M+U)}   & \textbf{30.83}$^\dagger$ & \textbf{34.34} & 1.5 & \textbf{37.21} & 33.38 & 1.6\\ 
         \cmidrule(lr){1-7}
            ~+~all \scriptsize{(P+R+M+U)}       & 31.92$^{\dagger\ddagger}$  & 35.48$^{\dagger\ddagger}$   & 4.8  & 36.65 & 32.88 & 4.5\\
         \bottomrule
    \end{tabular}
    \caption{Retrieval effectiveness of BM25 and BM25+RM3 using various indexing configurations. We also report the average number of keyphrases (\#kp). $\dagger$ and $\ddagger$ indicate significance over title \& abstract indexing and \underline{P}resent, respectively.}
    \label{tab:retrieval_results}
\end{table*}

\subsection{Distribution of gold-standard keyphrases under the PRMU scheme}
\label{sec:dist}

Table~\ref{tab:dist} shows the proportion of gold-standard, author-assigned keyphrases for each category in the different datasets. 
%
We also report results for the KP20k dataset~\cite{meng-etal-2017-deep}, which is used as training data by most neural keyphrase generation models.
%
%
We observe very similar distributions across datasets, with absent keyphrases accounting for about 40\% of the total number of keyphrases.
Interestingly, most of the absent keyphrases belong to the mixed and unseen categories, and therefore should provide some form of semantic expansion.
To have a precise idea of how many new words are actually added when indexing absent keyphrases, we compute the ratio (\%uw) of unique words from keyphrases that do not occur in their corresponding documents.
We find that only about 20\% of the words included in keyphrases contribute to expanding documents.
This surprisingly low percentage indicates that absent keyphrases play a much smaller role on document expansion than previously thought.
Yet, as we will see next, this small fraction of new words is behind much of the gains observed in retrieval effectiveness.

\subsection{Effect of indexing PRMU keyphrases on retrieval effectiveness}
\label{sec:ir}

Table~\ref{tab:retrieval_results} presents the results of retrieval models on documents supplemented with keyphrases from PRMU categories.
We see that adding keyphrases systematically improves retrieval effectiveness on both datasets, but a closer look reveals that the largest gains are obtained with Mixed and Unseen keyphrases.
This observation, combined with the fact that the number of Mixed and Unseen keyphrases is comparatively small (less than one on average), demonstrate that expanding documents is more effective than highlighting salient phrases for improving document retrieval performance.
The higher scores achieved when combining Mixed and Unseen keyphrases, compared to when combining Present and Reordered keyphrases, further confirm this conclusion.
Surprisingly, coupling query expansion (+RM3) with appending keyphrases yields conflicting results, which we attribute to the narrow set of topics (all related to IR) in ACM-CR that limits the vocabulary mismatch problem and makes it sensitive to semantic drift.
Another reason may be the incomplete nature of the relevance judgments, i.e.~that do not include uncited, yet relevant documents.
Here, the use of a co-cited probability metric as in~\cite{10.1145/2600428.2609585} may bring some new insights.

\begin{table}[!ht]
    \centering
    \begin{tabular}{l|rrrr|r}
    \toprule
    \textbf{Model} & \textbf{\%P} & \textbf{\%R} & \textbf{\%M} & \textbf{\%U} & \textbf{F@5} \\
    \midrule
        s2s+copy & 96.9 &  1.3 &  0.9 &  0.9 & 24.0\\
        s2s+corr & 89.7 &  7.1 &  2.5 &  0.8 & 22.1\\
    \bottomrule
    \end{tabular}
    \caption{Proportion of \underline{P}resent, \underline{R}eordered, \underline{M}ixed and \underline{U}nseen at the top-5 keyphrases on NTCIR-2. The f-measure against gold standard is also reported (F@5).}
    \label{tab:dist-model-ntcir-2}
\end{table}

\subsection{Analysis of keyphrase generation outputs under the PRMU scheme}
\label{sec:kg}

In this last experiment, we explore how the proposed categories fit into the outputs of neural keyphrase generation models.
Table~\ref{tab:dist-model-ntcir-2} shows the distributions over PRMU categories for two strong baseline models: \emph{s2s+copy}, a sequence-to-sequence model with attention and copying mechanisms~\cite{meng-etal-2017-deep}, and \emph{s2s+corr} which extends the aforementioned model with a coverage mechanism~\cite{chen-etal-2018-keyphrase}.
We observe that the output distributions are heavily skewed towards the Present category, indicating that the models have trouble producing keyphrases made up of new words.
Accordingly, the overall performance of these models is quite poor (about 20\% in f-measure), and mainly capped by the number of present keyphrases in the gold standard.
This advocates for more focus on training generative models to expand documents, rather than to imitate author-assigned annotation.


%

%
%



\section{Related Work}

Until recently, most previous models for predicting keyphrases were doing so by extracting the most salient noun phrases from documents~\cite{hasan-ng-2014-automatic}.
Keyphrase extraction models are usually divided into supervised models that cast keyphrase extraction either as a binary classification problem~\cite{10.1023/A:1009976227802,10.1145/313238.313437,hulth-2003-improved,10.1007/978-3-540-77094-7_41,medelyan-etal-2009-human,sterckx-etal-2016-supervised} or as a sequence labelling problem~\cite{augenstein-etal-2017-semeval,xiong-etal-2019-open,10.1145/3308558.3313642}, and unsupervised models that rely predominantly on graph-based ranking approaches~\cite{mihalcea-tarau-2004-textrank,litvak-last-2008-graph,wan-xiao-2008-collabrank,bougouin-etal-2013-topicrank,tixier-etal-2016-graph,boudin-2018-unsupervised}.
Note that none of these models can produce absent keyphrases.

A related line of research focuses on keyphrase assignment, that is, the task of selecting entries  from a predefined list of keyphrases (i.e.~a controlled vocabulary)~\cite{10.5555/254604.254610,10.1145/288627.288651,10.1145/1141753.1141819}.
Here, predicting keyphrases is treated as a multi-class classification task, and models can produce both present and absent keyphrases.
Further in that direction is~\cite{bougouin-etal-2016-keyphrase} that jointly performs keyphrase extraction and assignment using an unsupervised graph-based ranking model.

Also closely related to our work is previous research on document expansion~\cite{tao-etal-2006-language,10.1145/2348283.2348405}, and particularly recent work on supplementing document indexing with automatically generated queries~\cite{nogueira2019document,nogueira2019doc2query}.
These latter models augment texts with potential queries that, just as keyphrases, mitigate vocabulary mismatch and re-weight existing terms~\cite{lin2020pretrained}.
On the term weighting side, recent work shows that deep neural language models, in this case BERT~\cite{devlin-etal-2019-bert}, can be successfully applied to estimate document-specific term weights~\cite{10.1145/3397271.3401204}.

\section{Conclusion}

In this paper, we investigated the usefulness of absent keyphrases for document retrieval.
We showed that the commonly accepted definition of absent keyphrases is not sufficiently explicit in the context of IR, and proposed a finer-grained categorization scheme that allows for a better understanding of their impact on retrieval effectiveness.
Our code and data are publicly available at \url{https://github.com/boudinfl/redefining-absent-keyphrases}.

\section*{Acknowledgements}

We thank the reviewers for their valuable comments.
This work was supported by the French National Research Agency (ANR) through the DELICES project (ANR-19-CE38-0005-01).

\bibliography{anthology,biblio}

\begin{thebibliography}{52}
\expandafter\ifx\csname natexlab\endcsname\relax\def\natexlab#1{#1}\fi

\bibitem[{Abdul-Jaleel et~al.(2004)Abdul-Jaleel, Allan, Croft, Diaz, Larkey,
  Li, Smucker, and Wade}]{abdul2004umass}
Nasreen Abdul-Jaleel, James Allan, W~Bruce Croft, Fernando Diaz, Leah Larkey,
  Xiaoyan Li, Mark~D Smucker, and Courtney Wade. 2004.
\newblock Umass at trec 2004: Novelty and hard.
\newblock \emph{Computer Science Department Faculty Publication Series}, page
  189.

\bibitem[{Alzaidy et~al.(2019)Alzaidy, Caragea, and
  Giles}]{10.1145/3308558.3313642}
Rabah Alzaidy, Cornelia Caragea, and C.~Lee Giles. 2019.
\newblock \href {https://doi.org/10.1145/3308558.3313642} {Bi-lstm-crf sequence
  labeling for keyphrase extraction from scholarly documents}.
\newblock In \emph{The World Wide Web Conference}, WWW '19, page 2551–2557,
  New York, NY, USA. Association for Computing Machinery.

\bibitem[{Augenstein et~al.(2017)Augenstein, Das, Riedel, Vikraman, and
  McCallum}]{augenstein-etal-2017-semeval}
Isabelle Augenstein, Mrinal Das, Sebastian Riedel, Lakshmi Vikraman, and Andrew
  McCallum. 2017.
\newblock \href {https://doi.org/10.18653/v1/S17-2091} {{S}em{E}val 2017 task
  10: {S}cience{IE} - extracting keyphrases and relations from scientific
  publications}.
\newblock In \emph{Proceedings of the 11th International Workshop on Semantic
  Evaluation ({S}em{E}val-2017)}, pages 546--555, Vancouver, Canada.
  Association for Computational Linguistics.

\bibitem[{Bahuleyan and El~Asri(2020)}]{bahuleyan-el-asri-2020-diverse}
Hareesh Bahuleyan and Layla El~Asri. 2020.
\newblock \href {https://doi.org/10.18653/v1/2020.coling-main.462} {Diverse
  keyphrase generation with neural unlikelihood training}.
\newblock In \emph{Proceedings of the 28th International Conference on
  Computational Linguistics}, pages 5271--5287, Barcelona, Spain (Online).
  International Committee on Computational Linguistics.

\bibitem[{Boudin(2018)}]{boudin-2018-unsupervised}
Florian Boudin. 2018.
\newblock \href {https://doi.org/10.18653/v1/N18-2105} {Unsupervised keyphrase
  extraction with multipartite graphs}.
\newblock In \emph{Proceedings of the 2018 Conference of the North {A}merican
  Chapter of the Association for Computational Linguistics: Human Language
  Technologies, Volume 2 (Short Papers)}, pages 667--672, New Orleans,
  Louisiana. Association for Computational Linguistics.

\bibitem[{Boudin et~al.(2020)Boudin, Gallina, and
  Aizawa}]{boudin-etal-2020-keyphrase}
Florian Boudin, Ygor Gallina, and Akiko Aizawa. 2020.
\newblock \href {https://doi.org/10.18653/v1/2020.acl-main.105} {Keyphrase
  generation for scientific document retrieval}.
\newblock In \emph{Proceedings of the 58th Annual Meeting of the Association
  for Computational Linguistics}, pages 1118--1126, Online. Association for
  Computational Linguistics.

\bibitem[{Bougouin et~al.(2013)Bougouin, Boudin, and
  Daille}]{bougouin-etal-2013-topicrank}
Adrien Bougouin, Florian Boudin, and B{\'e}atrice Daille. 2013.
\newblock \href {https://www.aclweb.org/anthology/I13-1062} {{T}opic{R}ank:
  Graph-based topic ranking for keyphrase extraction}.
\newblock In \emph{Proceedings of the Sixth International Joint Conference on
  Natural Language Processing}, pages 543--551, Nagoya, Japan. Asian Federation
  of Natural Language Processing.

\bibitem[{Bougouin et~al.(2016)Bougouin, Boudin, and
  Daille}]{bougouin-etal-2016-keyphrase}
Adrien Bougouin, Florian Boudin, and B{\'e}atrice Daille. 2016.
\newblock \href {https://www.aclweb.org/anthology/C16-1277} {Keyphrase
  annotation with graph co-ranking}.
\newblock In \emph{Proceedings of {COLING} 2016, the 26th International
  Conference on Computational Linguistics: Technical Papers}, pages 2945--2955,
  Osaka, Japan. The COLING 2016 Organizing Committee.

\bibitem[{Chen et~al.(2018)Chen, Zhang, Wu, Yan, and
  Li}]{chen-etal-2018-keyphrase}
Jun Chen, Xiaoming Zhang, Yu~Wu, Zhao Yan, and Zhoujun Li. 2018.
\newblock \href {https://doi.org/10.18653/v1/D18-1439} {Keyphrase generation
  with correlation constraints}.
\newblock In \emph{Proceedings of the 2018 Conference on Empirical Methods in
  Natural Language Processing}, pages 4057--4066, Brussels, Belgium.
  Association for Computational Linguistics.

\bibitem[{Chen et~al.(2020)Chen, Chan, Li, and King}]{chen-etal-2020-exclusive}
Wang Chen, Hou~Pong Chan, Piji Li, and Irwin King. 2020.
\newblock \href {https://doi.org/10.18653/v1/2020.acl-main.103} {Exclusive
  hierarchical decoding for deep keyphrase generation}.
\newblock In \emph{Proceedings of the 58th Annual Meeting of the Association
  for Computational Linguistics}, pages 1095--1105, Online. Association for
  Computational Linguistics.

\bibitem[{Dai and Callan(2020)}]{10.1145/3397271.3401204}
Zhuyun Dai and Jamie Callan. 2020.
\newblock \href {https://doi.org/10.1145/3397271.3401204} {Context-aware term
  weighting for first stage passage retrieval}.
\newblock In \emph{Proceedings of the 43rd International ACM SIGIR Conference
  on Research and Development in Information Retrieval}, SIGIR '20, page
  1533–1536, New York, NY, USA. Association for Computing Machinery.

\bibitem[{Devlin et~al.(2019)Devlin, Chang, Lee, and
  Toutanova}]{devlin-etal-2019-bert}
Jacob Devlin, Ming-Wei Chang, Kenton Lee, and Kristina Toutanova. 2019.
\newblock \href {https://doi.org/10.18653/v1/N19-1423} {{BERT}: Pre-training of
  deep bidirectional transformers for language understanding}.
\newblock In \emph{Proceedings of the 2019 Conference of the North {A}merican
  Chapter of the Association for Computational Linguistics: Human Language
  Technologies, Volume 1 (Long and Short Papers)}, pages 4171--4186,
  Minneapolis, Minnesota. Association for Computational Linguistics.

\bibitem[{Dumais et~al.(1998)Dumais, Platt, Heckerman, and
  Sahami}]{10.1145/288627.288651}
Susan Dumais, John Platt, David Heckerman, and Mehran Sahami. 1998.
\newblock \href {https://doi.org/10.1145/288627.288651} {Inductive learning
  algorithms and representations for text categorization}.
\newblock In \emph{Proceedings of the Seventh International Conference on
  Information and Knowledge Management}, CIKM '98, page 148–155, New York,
  NY, USA. Association for Computing Machinery.

\bibitem[{Efron et~al.(2012)Efron, Organisciak, and
  Fenlon}]{10.1145/2348283.2348405}
Miles Efron, Peter Organisciak, and Katrina Fenlon. 2012.
\newblock \href {https://doi.org/10.1145/2348283.2348405} {Improving retrieval
  of short texts through document expansion}.
\newblock In \emph{Proceedings of the 35th International ACM SIGIR Conference
  on Research and Development in Information Retrieval}, SIGIR '12, page
  911–920, New York, NY, USA. Association for Computing Machinery.

\bibitem[{F{\"a}rber and Jatowt(2020)}]{farber2020citation}
Michael F{\"a}rber and Adam Jatowt. 2020.
\newblock Citation recommendation: Approaches and datasets.
\newblock \emph{arXiv preprint arXiv:2002.06961}.

\bibitem[{Florescu and Caragea(2017)}]{florescu-caragea-2017-positionrank}
Corina Florescu and Cornelia Caragea. 2017.
\newblock \href {https://doi.org/10.18653/v1/P17-1102} {{P}osition{R}ank: An
  unsupervised approach to keyphrase extraction from scholarly documents}.
\newblock In \emph{Proceedings of the 55th Annual Meeting of the Association
  for Computational Linguistics (Volume 1: Long Papers)}, pages 1105--1115,
  Vancouver, Canada. Association for Computational Linguistics.

\bibitem[{Furnas et~al.(1987)Furnas, Landauer, Gomez, and
  Dumais}]{10.1145/32206.32212}
G.~W. Furnas, T.~K. Landauer, L.~M. Gomez, and S.~T. Dumais. 1987.
\newblock \href {https://doi.org/10.1145/32206.32212} {The vocabulary problem
  in human-system communication}.
\newblock \emph{Commun. ACM}, 30(11):964–971.

\bibitem[{Gallina et~al.(2020)Gallina, Boudin, and
  Daille}]{10.1145/3383583.3398517}
Ygor Gallina, Florian Boudin, and B\'{e}atrice Daille. 2020.
\newblock \href {https://doi.org/10.1145/3383583.3398517} {Large-scale
  evaluation of keyphrase extraction models}.
\newblock In \emph{Proceedings of the ACM/IEEE Joint Conference on Digital
  Libraries in 2020}, JCDL '20, page 271–278, New York, NY, USA. Association
  for Computing Machinery.

\bibitem[{Gutwin et~al.(1999)Gutwin, Paynter, Witten, Nevill-Manning, and
  Frank}]{10.5555/338985.338996}
Carl Gutwin, Gordon Paynter, Ian Witten, Craig Nevill-Manning, and Eibe Frank.
  1999.
\newblock Improving browsing in digital libraries with keyphrase indexes.
\newblock \emph{Decis. Support Syst.}, 27(1–2):81–104.

\bibitem[{Hasan and Ng(2014)}]{hasan-ng-2014-automatic}
Kazi~Saidul Hasan and Vincent Ng. 2014.
\newblock \href {https://doi.org/10.3115/v1/P14-1119} {Automatic keyphrase
  extraction: A survey of the state of the art}.
\newblock In \emph{Proceedings of the 52nd Annual Meeting of the Association
  for Computational Linguistics (Volume 1: Long Papers)}, pages 1262--1273,
  Baltimore, Maryland. Association for Computational Linguistics.

\bibitem[{He et~al.(2010)He, Pei, Kifer, Mitra, and
  Giles}]{10.1145/1772690.1772734}
Qi~He, Jian Pei, Daniel Kifer, Prasenjit Mitra, and Lee Giles. 2010.
\newblock \href {https://doi.org/10.1145/1772690.1772734} {Context-aware
  citation recommendation}.
\newblock In \emph{Proceedings of the 19th International Conference on World
  Wide Web}, WWW '10, page 421–430, New York, NY, USA. Association for
  Computing Machinery.

\bibitem[{Huang et~al.(2019)Huang, Casey, G\l{}owacka, and
  Medlar}]{10.1145/3295750.3298953}
Chien-yu Huang, Arlene Casey, Dorota G\l{}owacka, and Alan Medlar. 2019.
\newblock \href {https://doi.org/10.1145/3295750.3298953} {Holes in the
  outline: Subject-dependent abstract quality and its implications for
  scientific literature search}.
\newblock In \emph{Proceedings of the 2019 Conference on Human Information
  Interaction and Retrieval}, CHIIR '19, page 289–293, New York, NY, USA.
  Association for Computing Machinery.

\bibitem[{Hulth(2003)}]{hulth-2003-improved}
Anette Hulth. 2003.
\newblock \href {https://www.aclweb.org/anthology/W03-1028} {Improved automatic
  keyword extraction given more linguistic knowledge}.
\newblock In \emph{Proceedings of the 2003 Conference on Empirical Methods in
  Natural Language Processing}, pages 216--223.

\bibitem[{Kando(2001)}]{ntcir-2}
Noriko Kando. 2001.
\newblock Overview of the second ntcir workshop.
\newblock In \emph{Proceedings of the Second NTCIR Workshop on Research in
  Chinese \& Japanese Text Retrieval and Text Summarization}.

\bibitem[{Leung and Kan(1997)}]{10.5555/254604.254610}
Chi-Hong Leung and Wing-Kay Kan. 1997.
\newblock A statistical learning approach to automatic indexing of controlled
  index terms.
\newblock \emph{J. Am. Soc. Inf. Sci.}, 48(1):55–66.

\bibitem[{Lin(2019)}]{Lin:2019:NHC:3308774.3308781}
Jimmy Lin. 2019.
\newblock \href {https://doi.org/10.1145/3308774.3308781} {The neural hype and
  comparisons against weak baselines}.
\newblock \emph{SIGIR Forum}, 52(2):40--51.

\bibitem[{Lin et~al.(2020)Lin, Nogueira, and Yates}]{lin2020pretrained}
Jimmy Lin, Rodrigo Nogueira, and Andrew Yates. 2020.
\newblock \href {http://arxiv.org/abs/2010.06467} {Pretrained transformers for
  text ranking: Bert and beyond}.

\bibitem[{Litvak and Last(2008)}]{litvak-last-2008-graph}
Marina Litvak and Mark Last. 2008.
\newblock \href {https://www.aclweb.org/anthology/W08-1404} {Graph-based
  keyword extraction for single-document summarization}.
\newblock In \emph{Coling 2008: Proceedings of the workshop Multi-source
  Multilingual Information Extraction and Summarization}, pages 17--24,
  Manchester, UK. Coling 2008 Organizing Committee.

\bibitem[{Livne et~al.(2014)Livne, Gokuladas, Teevan, Dumais, and
  Adar}]{10.1145/2600428.2609585}
Avishay Livne, Vivek Gokuladas, Jaime Teevan, Susan~T. Dumais, and Eytan Adar.
  2014.
\newblock \href {https://doi.org/10.1145/2600428.2609585} {Citesight:
  Supporting contextual citation recommendation using differential search}.
\newblock In \emph{Proceedings of the 37th International ACM SIGIR Conference
  on Research and Development in Information Retrieval}, SIGIR '14, page
  807–816, New York, NY, USA. Association for Computing Machinery.

\bibitem[{Medelyan et~al.(2009)Medelyan, Frank, and
  Witten}]{medelyan-etal-2009-human}
Olena Medelyan, Eibe Frank, and Ian~H. Witten. 2009.
\newblock \href {https://www.aclweb.org/anthology/D09-1137} {Human-competitive
  tagging using automatic keyphrase extraction}.
\newblock In \emph{Proceedings of the 2009 Conference on Empirical Methods in
  Natural Language Processing}, pages 1318--1327, Singapore. Association for
  Computational Linguistics.

\bibitem[{Medelyan and Witten(2006)}]{10.1145/1141753.1141819}
Olena Medelyan and Ian~H. Witten. 2006.
\newblock \href {https://doi.org/10.1145/1141753.1141819} {Thesaurus based
  automatic keyphrase indexing}.
\newblock In \emph{Proceedings of the 6th ACM/IEEE-CS Joint Conference on
  Digital Libraries}, JCDL '06, page 296–297, New York, NY, USA. Association
  for Computing Machinery.

\bibitem[{Meng et~al.(2017)Meng, Zhao, Han, He, Brusilovsky, and
  Chi}]{meng-etal-2017-deep}
Rui Meng, Sanqiang Zhao, Shuguang Han, Daqing He, Peter Brusilovsky, and
  Yu~Chi. 2017.
\newblock \href {https://doi.org/10.18653/v1/P17-1054} {Deep keyphrase
  generation}.
\newblock In \emph{Proceedings of the 55th Annual Meeting of the Association
  for Computational Linguistics (Volume 1: Long Papers)}, pages 582--592,
  Vancouver, Canada. Association for Computational Linguistics.

\bibitem[{Mihalcea and Tarau(2004)}]{mihalcea-tarau-2004-textrank}
Rada Mihalcea and Paul Tarau. 2004.
\newblock \href {https://www.aclweb.org/anthology/W04-3252} {{T}ext{R}ank:
  Bringing order into text}.
\newblock In \emph{Proceedings of the 2004 Conference on Empirical Methods in
  Natural Language Processing}, pages 404--411, Barcelona, Spain. Association
  for Computational Linguistics.

\bibitem[{Nguyen and Kan(2007)}]{10.1007/978-3-540-77094-7_41}
Thuy~Dung Nguyen and Min-Yen Kan. 2007.
\newblock Keyphrase extraction in scientific publications.
\newblock In \emph{Asian Digital Libraries. Looking Back 10 Years and Forging
  New Frontiers}, pages 317--326, Berlin, Heidelberg. Springer Berlin
  Heidelberg.

\bibitem[{Nogueira and Lin(2019)}]{nogueira2019doc2query}
Rodrigo Nogueira and Jimmy Lin. 2019.
\newblock From doc2query to doctttttquery.

\bibitem[{Nogueira et~al.(2019)Nogueira, Yang, Lin, and
  Cho}]{nogueira2019document}
Rodrigo Nogueira, Wei Yang, Jimmy Lin, and Kyunghyun Cho. 2019.
\newblock \href {http://arxiv.org/abs/1904.08375} {Document expansion by query
  prediction}.

\bibitem[{Roy(2017)}]{10.1145/3132847.3133085}
Dwaipayan Roy. 2017.
\newblock \href {https://doi.org/10.1145/3132847.3133085} {An improved test
  collection and baselines for bibliographic citation recommendation}.
\newblock In \emph{Proceedings of the 2017 ACM on Conference on Information and
  Knowledge Management}, CIKM '17, page 2271–2274, New York, NY, USA.
  Association for Computing Machinery.

\bibitem[{Santosh et~al.(2020)Santosh, Kumar~Sanyal, Bhowmick, and
  Das}]{santosh-etal-2020-sasake}
T.y.s.s Santosh, Debarshi Kumar~Sanyal, Plaban~Kumar Bhowmick, and
  Partha~Pratim Das. 2020.
\newblock \href {https://doi.org/10.18653/v1/2020.coling-main.469} {{S}a{SAKE}:
  Syntax and semantics aware keyphrase extraction from research papers}.
\newblock In \emph{Proceedings of the 28th International Conference on
  Computational Linguistics}, pages 5372--5383, Barcelona, Spain (Online).
  International Committee on Computational Linguistics.

\bibitem[{Smucker et~al.(2007)Smucker, Allan, and
  Carterette}]{Smucker:2007:CSS:1321440.1321528}
Mark~D. Smucker, James Allan, and Ben Carterette. 2007.
\newblock \href {https://doi.org/10.1145/1321440.1321528} {A comparison of
  statistical significance tests for information retrieval evaluation}.
\newblock In \emph{Proceedings of the Sixteenth ACM Conference on Conference on
  Information and Knowledge Management}, CIKM '07, pages 623--632, New York,
  NY, USA. ACM.

\bibitem[{Sterckx et~al.(2016)Sterckx, Caragea, Demeester, and
  Develder}]{sterckx-etal-2016-supervised}
Lucas Sterckx, Cornelia Caragea, Thomas Demeester, and Chris Develder. 2016.
\newblock \href {https://doi.org/10.18653/v1/D16-1198} {Supervised keyphrase
  extraction as positive unlabeled learning}.
\newblock In \emph{Proceedings of the 2016 Conference on Empirical Methods in
  Natural Language Processing}, pages 1924--1929, Austin, Texas. Association
  for Computational Linguistics.

\bibitem[{Sun et~al.(2019)Sun, Tang, Du, Deng, and
  Nie}]{10.1145/3331184.3331219}
Zhiqing Sun, Jian Tang, Pan Du, Zhi-Hong Deng, and Jian-Yun Nie. 2019.
\newblock \href {https://doi.org/10.1145/3331184.3331219} {Divgraphpointer: A
  graph pointer network for extracting diverse keyphrases}.
\newblock In \emph{Proceedings of the 42nd International ACM SIGIR Conference
  on Research and Development in Information Retrieval}, SIGIR'19, page
  755–764, New York, NY, USA. Association for Computing Machinery.

\bibitem[{Tao et~al.(2006)Tao, Wang, Mei, and Zhai}]{tao-etal-2006-language}
Tao Tao, Xuanhui Wang, Qiaozhu Mei, and ChengXiang Zhai. 2006.
\newblock \href {https://www.aclweb.org/anthology/N06-1052} {Language model
  information retrieval with document expansion}.
\newblock In \emph{Proceedings of the Human Language Technology Conference of
  the {NAACL}, Main Conference}, pages 407--414, New York City, USA.
  Association for Computational Linguistics.

\bibitem[{Tixier et~al.(2016)Tixier, Malliaros, and
  Vazirgiannis}]{tixier-etal-2016-graph}
Antoine Tixier, Fragkiskos Malliaros, and Michalis Vazirgiannis. 2016.
\newblock \href {https://doi.org/10.18653/v1/D16-1191} {A graph
  degeneracy-based approach to keyword extraction}.
\newblock In \emph{Proceedings of the 2016 Conference on Empirical Methods in
  Natural Language Processing}, pages 1860--1870, Austin, Texas. Association
  for Computational Linguistics.

\bibitem[{Turney(2000)}]{10.1023/A:1009976227802}
Peter~D. Turney. 2000.
\newblock \href {https://doi.org/10.1023/A:1009976227802} {Learning algorithms
  for keyphrase extraction}.
\newblock \emph{Inf. Retr.}, 2(4):303–336.

\bibitem[{Wan and Xiao(2008)}]{wan-xiao-2008-collabrank}
Xiaojun Wan and Jianguo Xiao. 2008.
\newblock \href {https://www.aclweb.org/anthology/C08-1122} {{C}ollab{R}ank:
  Towards a collaborative approach to single-document keyphrase extraction}.
\newblock In \emph{Proceedings of the 22nd International Conference on
  Computational Linguistics (Coling 2008)}, pages 969--976, Manchester, UK.
  Coling 2008 Organizing Committee.

\bibitem[{Wang et~al.(2020)Wang, Fan, and Rose}]{wang-etal-2020-incorporating}
Yansen Wang, Zhen Fan, and Carolyn Rose. 2020.
\newblock \href {https://doi.org/10.18653/v1/2020.emnlp-main.140}
  {Incorporating multimodal information in open-domain web keyphrase
  extraction}.
\newblock In \emph{Proceedings of the 2020 Conference on Empirical Methods in
  Natural Language Processing (EMNLP)}, pages 1790--1800, Online. Association
  for Computational Linguistics.

\bibitem[{Witten et~al.(1999)Witten, Paynter, Frank, Gutwin, and
  Nevill-Manning}]{10.1145/313238.313437}
Ian~H. Witten, Gordon~W. Paynter, Eibe Frank, Carl Gutwin, and Craig~G.
  Nevill-Manning. 1999.
\newblock \href {https://doi.org/10.1145/313238.313437} {Kea: Practical
  automatic keyphrase extraction}.
\newblock In \emph{Proceedings of the Fourth ACM Conference on Digital
  Libraries}, DL '99, page 254–255, New York, NY, USA. Association for
  Computing Machinery.

\bibitem[{Xiong et~al.(2019)Xiong, Hu, Xiong, Campos, and
  Overwijk}]{xiong-etal-2019-open}
Lee Xiong, Chuan Hu, Chenyan Xiong, Daniel Campos, and Arnold Overwijk. 2019.
\newblock \href {https://doi.org/10.18653/v1/D19-1521} {Open domain web
  keyphrase extraction beyond language modeling}.
\newblock In \emph{Proceedings of the 2019 Conference on Empirical Methods in
  Natural Language Processing and the 9th International Joint Conference on
  Natural Language Processing (EMNLP-IJCNLP)}, pages 5175--5184, Hong Kong,
  China. Association for Computational Linguistics.

\bibitem[{Yang et~al.(2017)Yang, Fang, and Lin}]{Yang:2017:AEU:3077136.3080721}
Peilin Yang, Hui Fang, and Jimmy Lin. 2017.
\newblock \href {https://doi.org/10.1145/3077136.3080721} {Anserini: Enabling
  the use of lucene for information retrieval research}.
\newblock In \emph{Proceedings of the 40th International ACM SIGIR Conference
  on Research and Development in Information Retrieval}, SIGIR '17, pages
  1253--1256, New York, NY, USA. ACM.

\bibitem[{Yang et~al.(2019)Yang, Lu, Yang, and Lin}]{10.1145/3331184.3331340}
Wei Yang, Kuang Lu, Peilin Yang, and Jimmy Lin. 2019.
\newblock \href {https://doi.org/10.1145/3331184.3331340} {Critically examining
  the "neural hype": Weak baselines and the additivity of effectiveness gains
  from neural ranking models}.
\newblock In \emph{Proceedings of the 42nd International ACM SIGIR Conference
  on Research and Development in Information Retrieval}, SIGIR'19, page
  1129–1132, New York, NY, USA. Association for Computing Machinery.

\bibitem[{Zhai(1997)}]{zhai-1997-fast}
Chengxiang Zhai. 1997.
\newblock \href {https://doi.org/10.3115/974557.974603} {Fast statistical
  parsing of noun phrases for document indexing}.
\newblock In \emph{Fifth Conference on Applied Natural Language Processing},
  pages 312--319, Washington, DC, USA. Association for Computational
  Linguistics.

\bibitem[{Zhao and Zhang(2019)}]{zhao-zhang-2019-incorporating}
Jing Zhao and Yuxiang Zhang. 2019.
\newblock \href {https://doi.org/10.18653/v1/P19-1515} {Incorporating
  linguistic constraints into keyphrase generation}.
\newblock In \emph{Proceedings of the 57th Annual Meeting of the Association
  for Computational Linguistics}, pages 5224--5233, Florence, Italy.
  Association for Computational Linguistics.

\end{thebibliography}
\bibliographystyle{acl_natbib}

\appendix

\section{Computing fine-grained PRMU categories}
\label{sec:code}

The python code below (Figure~\ref{fig:finer_category_code}) showcases how PRMU keyphrase categories proposed in this paper are computed.
Note that \texttt{kw} and \texttt{doc} are preprocessed lists of tokens (lowercased and in stemmed forms), and that title is separated from the abstract using a special token in order to prevent the last word of the title to be contiguous with the first word of the abstract.

These functions are written to be an understandable reference, the code shared with this work uses a different implementation.

\begin{figure}[ht!]
    \centering
\begin{minted}{python}
def contains(small, big):
    # Checks whether `small` appear
    #  contiguously in `big`
    for i in range(
            len(big) - len(small) + 1):
        match_len = 0
        for j in range(len(small)):
            if big[i + j] == small[j]:
                match_len += 1
            else:
                break
        
        if match_len == len(small):
            # Every elements were found
            return True
    return False


def kw_category(kw, doc):
    if contains(kw, doc):
        return 'P'  # Present
    else:
        abs_words = [w for w in kw
                     if w not in doc]
        if len(abs_words) == 0:
            return 'R'  # Reordered
        elif len(abs_words) < len(kw):
            return 'M'  # Mixed
        elif len(abs_words) == len(kw):
            return 'U'  # Unseen
\end{minted}
    \caption{Python code computing the fine-grained PRMU category of a keyphrase (\texttt{kw}) with respect to a given document (\texttt{doc}).}
    \label{fig:finer_category_code}
\end{figure}

\end{document}